%% file: main.tex
\pdfoutput=1
\documentclass[sigconf]{acmart}

\AtBeginDocument{%
  }

\setcopyright{acmlicensed}
\copyrightyear{2018}
\acmYear{2018}
\acmDOI{XXXXXXX.XXXXXXX}

\usepackage{enumitem}
\usepackage{multirow}
\usepackage{bm}
\usepackage{subfigure}
\usepackage{textcomp}%

\acmConference[WWW'25]{The 2025 ACM Web Conference}{Apri 28--May 02,
  2025}{Sydney, Australia}

\acmISBN{978-1-4503-XXXX-X/18/06}

\begin{document}

\title{The 1st Workshop on Human-Centered Recommender Systems}
\subtitle{\url{https://human-centeredrec.github.io/}}

\author{Kaike Zhang}
\affiliation{%
  \institution{University of Chinese Academy of Sciences}
  \country{}
}

\email{kaikezhang99@gmail.com}

\author{Yunfan Wu}
\affiliation{%
  \institution{University of Chinese Academy of Sciences}
  \country{}
}
\email{wuyunfan666@gmail.com}

\author{Yougang Lyu}
\affiliation{%
  \institution{University of Amsterdam}
  \country{}
}
\email{youganglyu@gmail.com}

\author{Du Su}
\affiliation{%
  \institution{CAS Key Laboratory of AI Safety, Institute of Computing Technology, Chinese Academy of Sciences}
  \country{}
}
\email{sudu@ict.ac.cn}

\author{Yingqiang Ge}
\affiliation{%
  \institution{Amazon}
  \country{}
}
\email{gyq@amazon.com}

\author{Shuchang Liu}
\affiliation{%
  \institution{Kuaishou}
  \country{}
}
\email{liushuchang@kuaishou.com}

\author{Qi Cao}
\affiliation{%
  \institution{CAS Key Laboratory of AI Safety, Institute of Computing Technology, Chinese Academy of Sciences}
  \country{}
}
\email{caoqi@ict.ac.cn}

\author{Zhaochun Ren}
\affiliation{%
  \institution{Leiden University}
  \country{}
}
\email{z.ren@liacs.leidenuniv.nl}

\author{Fei Sun}
\affiliation{%
  \institution{CAS Key Laboratory of AI Safety, Institute of Computing Technology, Chinese Academy of Sciences}
  \country{}
}
\email{sunfei@ict.ac.cn}

\renewcommand{\shortauthors}{Kaike Zhang et al.}

\begin{abstract}

\input{Section/0-Abstract}
\end{abstract}



\maketitle

\section{INTRODUCTION}

\input{Section/1-Introduction}

\section{SCOPE AND TOPICS}
\label{sec:scope}
\input{Section/1-Scope}

\section{RATIONALE}

\input{Section/2-Rationale}

\section{WORKSHOP PROGRAM FORMAT}
\input{Section/3-Format}

\section{CALL FOR PAPERS}
\input{Section/4-Call}

\section{ORGANIZERS}

\input{Section/5-Organizer}

\bibliographystyle{ACM-Reference-Format}
\bibliography{ref}

\end{document}

%% file: Section/0-Abstract.tex
Recommender systems are quintessential applications of human-computer interaction. Widely utilized in daily life, they offer significant convenience but also present numerous challenges, such as the information cocoon effect, privacy concerns, fairness issues, and more. Consequently, this workshop aims to provide a platform for researchers to explore the development of Human-Centered Recommender Systems~(HCRS). HCRS refers to the creation of recommender systems that prioritize human needs, values, and capabilities at the core of their design and operation. In this workshop, topics will include, but are not limited to, robustness, privacy, transparency, fairness, diversity, accountability, ethical considerations, and user-friendly design. We hope to engage in discussions on how to implement and enhance these properties in recommender systems. Additionally, participants will explore diverse evaluation methods, including innovative metrics that capture user satisfaction and trust. This workshop seeks to foster a collaborative environment for researchers to share insights and advance the field toward more ethical, user-centric, and socially responsible recommender systems. 


%% file: Section/1-Introduction.tex
Recommender systems, as essential tools for managing the exponential growth of information available online, influence users' decisions in multifaceted ways across domains such as e-commerce, media consumption, and social networking. However, they also present challenges, including the information cocoon effect, privacy issues, and fairness concerns. Recently, there has been a growing body of work emphasizing the importance of developing Human-Centered Recommender Systems~(HCRS)~\cite{konstan2021human, silva2024leveraging}. HCRS refers to the development of recommender systems that prioritize human needs, values, and capabilities at the core of their design and operation.

While concepts like Trustworthy Recommender Systems and Responsible Recommender Systems share similarities with HCRS in addressing ethical and societal concerns, they differ in their primary focus. Trustworthy Recommender Systems aim to build systems that users can rely on, emphasizing aspects such as transparency, robustness, privacy, and security~\cite{ge2022survey, wang2024trustworthy}. Responsible Recommender Systems, on the other hand, concentrate on aligning system operations with ethical principles and social responsibilities, often focusing on fairness, accountability, and mitigating biases~\cite{kazienko2024toward}. 

In contrast, Human-Centered Recommender Systems place the human user at the center of the design process, striving to enhance user well-being, satisfaction, and empowerment by deeply considering human values, needs, and experiences. HCRS not only addresses trustworthiness and responsibility but also actively involves users in the design and evaluation of recommender systems to ensure they align with users' goals and capabilities~\cite{konstan2021human, shneiderman2020human, lowdermilk2013user}. This involves methodologies such as participatory design, where users collaborate in the creation process, and user studies that inform system development with real-world insights.

%% file: Section/1-Scope.tex
This workshop offers researchers an invaluable collaborative platform to present the latest advancements in the rapidly evolving field of HCRS. By fostering interdisciplinary dialogue and knowledge exchange, we aim to advance the understanding and implementation of recommender systems that are not only technically effective but also ethically sound, user-centered, and socially responsible. We welcome original submissions focusing on:

\begin{itemize}[leftmargin=*]
    \item \textbf{Robustness}: Fraudster Detection, Defense Against Adversarial Attacks, Vulnerabilities of LLM-based RS, Certifiable Robustness, Denoising, Data Sparsity, Cross-Domain Robustness, User-Aware Robustness, etc.

    \item \textbf{Privacy}: Differential Privacy, Federated Learning in RS, Data Ownership, Privacy Risks in LLM-based RS, Data Anonymization, Membership Inference Attack, Data Minimization, Unlearning, etc.

    \item \textbf{Transparency}: Explainable RS, Interpretable RS, User-Centric Explanation Generation, Causal Explanations, Neurosymbolic Reasoning for RS, LLMs for Transparent RS, etc.

    \item \textbf{Fairness and Bias}: Debiasing in RS, Fairness and Bias in LLM-based RS, etc.

    \item \textbf{Diversity}: Content Diversity, Recommendation Diversity, User Perceptions of Diversity and Personalization, Addressing Filter Bubbles and Echo Chambers, etc.

    \item \textbf{Ethics of Recommender Systems}: Ethical Frameworks, Mitigation of Misinformation Spread, Ethical Implications of Personalized Content, User Consent and Ethical Data Usage, Ethical Challenges in LLM-based RS, etc.

    \item \textbf{Accountability}: Traceability, Responsible RS, Controllable Recommendations, etc.

    \item \textbf{Human-Computer Interaction Design}: User Interface Design, Interactive and Conversational RS, Accessibility and Inclusivity, etc.

    \item \textbf{Evaluation, Auditing, and Governance}: Evaluation Metrics, User Studies, Algorithm Auditing, Simulation, Governance Models, etc.
\end{itemize}

%% file: Section/2-Rationale.tex
The rapid advancement of human-centered recommender systems has catalyzed a wave of innovative research efforts in recent months~\cite{zhang2023robust, zhang2024lorec, zhang2024improving, wu2024accelerating, wang2024trustworthy, deldjoo2024fairness, liu2024privacy, yoo2024ensuring, balloccu2024explainable, zhou2024bee, zhang2024understanding}. Consequently, hosting this workshop is essential for fostering discussions on these evolving directions in the field.

\subsection{Relevance}
This workshop, centered on human-centered recommender systems, aligns seamlessly with the Web conference, highlighting a pivotal research trend within recommender systems—an area of critical importance at the conference. Furthermore, recommender systems are fundamentally designed to assist users in filtering information. By emphasizing human-centered concepts, we aim to enhance user satisfaction and acceptance of these systems, ultimately enabling them to provide more effective personalized suggestions for social good and further advancing the field.

\subsection{Objectives and Expected Outcomes}
This workshop encourages researchers to propose new theoretical frameworks, interdisciplinary approaches, and perspectives for human-centered recommender systems. We will also advocate for the adoption of advanced technologies, such as large language models, to enhance the human-centric qualities of existing recommender systems. Additionally, participants will be encouraged to develop innovative evaluation frameworks and metrics tailored to assessing these qualities. In the long term, this research focus is expected to drive the evolution of recommender systems into broader domains, fostering their applicability across various contexts and contributing positively to the well-being of society.

\subsection{Target Audience}
The appeal of this workshop lies in its commitment to the continuously evolving landscape of recommender systems. It aims to attract a diverse audience, including researchers, industry experts, and academics. The workshop will provide these stakeholders with a unique forum to share innovative ideas, methodologies, and achievements, fostering interdisciplinary collaboration and exploring new applications. By bringing together a broad spectrum of participants, we hope to stimulate rich discussions and catalyze the advancement of human-centered approaches in the realm of recommender systems.

\subsection{Diversity and Inclusion}

\textbf{Diversity among Organizers}. Our organizing team is characterized by diversity in terms of gender, affiliations, countries of origin, scientific backgrounds, and levels of seniority. This diverse composition enhances our capacity to consider a wide range of perspectives, ultimately enriching the discussions and outcomes of the workshop.

\textbf{Diversity among Speakers}. We will invite researchers with varied academic backgrounds to deliver keynote talks, ensuring a broad spectrum of insights and expertise. Additionally, we encourage submissions from scholars worldwide, aiming to showcase the work of diverse voices accepted into this workshop.

To promote diversity among our invited panel speakers, we will curate a mix of participants, including academics and industry professionals, representing various levels of seniority and experience in their respective fields. This approach will foster rich dialogue, allowing for a comprehensive exchange of ideas and perspectives that reflect the multifaceted nature of human-centered recommender systems.

%% file: Section/3-Format.tex
This workshop will be conducted over a half-day session. We will invite two researchers in the field to deliver 30-minute keynote talks, providing valuable insights into the latest advancements and future prospects of human-centered recommender systems. Additionally, we will host a panel discussion featuring several senior researchers and developers, focusing on future directions and challenges in this domain. 

We expect to accept 8-10 papers at this workshop. We will also incorporate a paper presentation session to allow participants to share their research findings and foster academic discourse. The preliminary program schedule is presented in Table~\ref{tab:schedule}.\footnote{Please note that the paper acceptance schedule and program timeline are provisional and may be subject to adjustments based on the conference chair's requirements.}

\begin{table}[t]
  \centering
    \caption{Program Schedule}
    \resizebox{0.4\textwidth}{!}{
\begin{tabular}{lc}
    \toprule
    \textbf{Event} & \textbf{Time} \\
    \midrule
     Opening Remarks from Co-Chairs & 08:40–08:50 \\
     Keynote Talk \#1 followed by Q\&A & 08:50–09:20 \\
     Paper Session \#1 & 09:20–10:00 \\
     Tea Break & 10:00–10:30 \\
     Keynote Talk \#2 followed by Q\&A & 10:30–11:00 \\
     Paper Session \#2 & 11:00–11:40 \\
     Panel Discussion \& Closing Remarks & 11:40–12:00 \\
    \bottomrule
    \end{tabular}
    }
  \label{tab:schedule}%
\end{table}

This structured schedule is designed to facilitate dynamic interactions among participants, encouraging networking opportunities and collaborative discussions. We anticipate that the combination of keynote addresses, research presentations, and a panel discussion will provide a comprehensive exploration of human-centered recommender systems, fostering an environment conducive to innovation and the exchange of ideas.

We have contacted the following individuals to serve on our Program Committee: Yuan Zhang (ByteDance), Xinyu Lin (National University of Singapore), Chen Xu (Renmin University of China), Yuanhao Liu (Institute of Computing Technology), Xiao Lin (Kuaishou), Jiakai Tang (Renmin University of China), Huizhong Guo (Zhejiang University), Yuyue Zhao (University of Science and Technology of China), Zhiyu He (Tsinghua University) and Zhaolin Gao (Cornel University). We also plan to invite the following individuals, whom we have not yet contacted: Allegra De Filippo (University of Bologna), Fabrizio Silvestri (Sapienza University of Rome), Michael D. Ekstrand (Drexel University), Anshuman Chhabra (University of South Florida), Ivan Srba (Kempelen Institute of Intelligent Technologies).


%% file: Section/4-Call.tex
This workshop aims to encourage innovative research on human-centered recommender systems, particularly in the areas of robustness, privacy, transparency, fairness, diversity, accountability, ethical considerations, user-friendly design, and evaluation methods. We summarize the detailed objectives and scope in Section~\ref{sec:scope}.

All submitted papers must be formatted as a single PDF document according to the ACM WWW 2025 template. Manuscripts may range from 4 to 8 pages in length, with unlimited pages allowed for references. Authors are encouraged to determine the appropriate length for their submissions, as there is no distinction made between long and short papers.

All submissions will adhere to a double-blind review policy, ensuring that both the authors and reviewers remain anonymous throughout the evaluation process. Each paper will undergo a rigorous review procedure, with expert peer reviewers assessing submissions based on their relevance to the workshop, scientific novelty, and technical quality.

The important dates for the submission process are as follows:
\begin{itemize}
    \item \textbf{Submission Deadlin}e: December 18, 2024
    \item \textbf{Paper Acceptance Notification}: January 13, 2025
    \item \textbf{Camera-Ready Submission}: February 2, 2025
\end{itemize}

%% file: Section/5-Organizer.tex
\begin{itemize}[leftmargin=*]
    \item \textbf{Kaike Zhang} \\
    \underline{Email}: kaikezhang99@gmail.com \\
    \underline{Affiliation}: University of Chinese Academy of Sciences \\
    \underline{Biography}: Kaike Zhang is a Ph.D. candidate at University of Chinese Academy of Sciences, under the supervision of Prof. Xueqi Cheng and Prof. Xinran Liu. His research focuses on trustworthy recommender systems. He has published in top-tier conferences such as NeurIPS, SIGIR, SIGKDD, and RecSys. He has also served as a reviewer and Program Committee member for high-level conferences and journals, including SIGIR, ICLR, WWW, and TKDD.
    
    \item \textbf{Yunfan Wu} \\
    \underline{Email}: wuyunfan666@gmail.com \\
    \underline{Affiliation}: University of Chinese Academy of Sciences \\
    \underline{Biography}: Yunfan Wu is a Ph.D. candidate at University of Chinese Academy of Sciences. His research focuses on trustworthy recommender systems. He has published in leading conferences and journals such as NeurIPS, SIGIR, CIKM, RecSys, and TKDE.
    
    \item \textbf{Yougang Lyu} \\
    \underline{Email}: youganglyu@gmail.com\\
    \underline{Affiliation}: University of Amsterdam \\
    \underline{Biography}: Yougang Lyu is a Ph.D. candidate at the University of Amsterdam, under the supervision of Prof. Maarten de Rijke and Dr. Zhaochun Ren. His research interests lie in LLM-based recommender systems and aligning LLMs with human values. He has published in top conferences and journals such as AAAI, RecSys, EMNLP, and IPM, and received the Best Paper Award at RecSys 2024. He has also served as a reviewer and Program Committee member for top conferences and journals, including SIGIR, ACL, EMNLP, CIKM, NAACL, COLING, and IPM.
    
    \item \textbf{Du Su} \\
    \underline{Email}: sudu@ict.ac.cn\\
    \underline{Affiliation}: CAS Key Laboratory of AI Safety, Institute of Computing Technology, Chinese Academy of Sciences \\
    \underline{Biography}: Dr. Du Su is an Assistant Professor at the CAS Key Laboratory of AI Safety, Institute of Computing Technology, Chinese Academy of Sciences. His research interests are in AI safety and security, with a specific focus on technologies for monitoring and assessing AI risks. His goal is to develop an agent-based sandbox environment for algorithm simulation, aimed at assessing algorithms' safety properties from the user's perspective, including privacy, fairness, and filter bubbles. He has published his work in top conferences such as KDD, WWW, and EMNLP.
    
    \item \textbf{Yingqiang Ge} \\
    \underline{Email}: gyq@amazon.com \\
    \underline{Affiliation}: Amazon \\
    \underline{Biography}: Dr. Yingqiang Ge is an Applied Scientist at Amazon. He earned his Ph.D. in Computer Science from Rutgers University under the supervision of Prof. Yongfeng Zhang. His research focuses on Trustworthy AI and Large Language Models in Recommender Systems. He has published in conferences such as WWW, SIGIR, KDD, NeurIPS, CIKM, WSDM, and TORS. Dr. Ge has co-organized tutorials at SIGIR 2021 and CIKM 2021, and has served as a Program Committee member for numerous top-tier computer science conferences and journals.
    
    \item \textbf{Shuchang Liu} \\
    \underline{Email}: liushuchang@kuaishou.com\\
    \underline{Affiliation}: Kuaishou \\
    \underline{Biography}: Dr. Shuchang Liu is an Applied Scientist at Kuaishou Technology, Beijing. He earned his Ph.D. in Computer Science from Rutgers University in 2022, under the supervision of Prof. Yongfeng Zhang. His research interests are in recommender systems. He has served as a Program Committee member for conferences including AAAI (2023-2025), IJCAI (2021-2024), KDD (2022-2025), RecSys (2024/2025), TOIS, TORS, and WWW (2021/2022).
    
    \item \textbf{Qi Cao} \\
    \underline{Email}: caoqi@ict.ac.cn\\
    \underline{Affiliation}: CAS Key Laboratory of AI Safety, Institute of Computing Technology, Chinese Academy of Sciences \\
    \underline{Biography}: Dr. Qi Cao is an Associate Professor at the Institute of Computing Technology, Chinese Academy of Sciences. Her research interests include social media analysis, robustness of recommender systems, and fairness auditing in machine learning. She has published over 30 papers in high-level conferences and journals such as KDD, WWW, NeurIPS, SIGIR, and WSDM. She was nominated for the Chinese Information Processing Society of China (CIPS) Outstanding Doctoral Dissertation Award. She has also served as a Program Committee member for conferences including AAAI, WWW, NeurIPS, and ICLR, and as a reviewer for journals such as TKDE, TOIS, and TKDD.
    
    \item \textbf{Zhaochun Ren} \\
    \underline{Email}: z.ren@liacs.leidenuniv.nl\\
    \underline{Affiliation}: Leiden University \\
    \underline{Biography}: Dr. Zhaochun Ren is an Associate Professor at Leiden University. His research interests focus on joint research in information retrieval and natural language processing, with an emphasis on conversational information-seeking, question-answering, and recommender systems. He aims to develop intelligent systems that can address complex user requests and solve core challenges in both fields. He has over two years of experience working on e-commerce search and recommendation at JD.com. He has co-organized workshops at SIGIR (2020, 2021), WSDM (2019, 2020), and ECIR (2024, 2025).
    
    \item \textbf{Fei Sun} \\
    \underline{Email}: sunfei@ict.ac.cn\\
    \underline{Affiliation}: CAS Key Laboratory of AI Safety, Institute of Computing Technology, Chinese Academy of Sciences \\
    \underline{Biography}: Dr. Fei Sun is an Associate Professor at the Institute of Computing Technology, Chinese Academy of Sciences. His research spans recommender systems and natural language processing, with a particular emphasis on safety-oriented topics such as privacy, fairness, and interpretability. He has published over 50 papers in leading conferences and journals, including WWW, SIGIR, and TOIS, and was awarded the Best Long Paper Runner-Up at RecSys 2019. Before joining academia, he spent five years at Alibaba focusing on e-commerce recommendation. He has also contributed to the community by co-organizing workshops, including one at WSDM 2020.
\end{itemize}

%% file: main.bbl

\begin{thebibliography}{17}


\ifx \showCODEN    \undefined \def \showCODEN     #1{\unskip}     \fi
\ifx \showDOI      \undefined \def \showDOI       #1{#1}\fi
\ifx \showISBNx    \undefined \def \showISBNx     #1{\unskip}     \fi
\ifx \showISBNxiii \undefined \def \showISBNxiii  #1{\unskip}     \fi
\ifx \showISSN     \undefined \def \showISSN      #1{\unskip}     \fi
\ifx \showLCCN     \undefined \def \showLCCN      #1{\unskip}     \fi
\ifx \shownote     \undefined \def \shownote      #1{#1}          \fi
\ifx \showarticletitle \undefined \def \showarticletitle #1{#1}   \fi
\ifx \showURL      \undefined \def \showURL       {\relax}        \fi
\providecommand\bibfield[2]{#2}
\providecommand\bibinfo[2]{#2}
\providecommand\natexlab[1]{#1}
\providecommand\showeprint[2][]{arXiv:#2}

\bibitem[Balloccu et~al\mbox{.}(2024)]%
        {balloccu2024explainable}
\bibfield{author}{\bibinfo{person}{Giacomo Balloccu}, \bibinfo{person}{Ludovico Boratto}, \bibinfo{person}{Gianni Fenu}, \bibinfo{person}{Francesca~Maridina Malloci}, {and} \bibinfo{person}{Mirko Marras}.} \bibinfo{year}{2024}\natexlab{}.
\newblock \showarticletitle{Explainable Recommender Systems with Knowledge Graphs and Language Models}. In \bibinfo{booktitle}{\emph{European Conference on Information Retrieval}}. Springer, \bibinfo{pages}{352--357}.
\newblock


\bibitem[Deldjoo et~al\mbox{.}(2024)]%
        {deldjoo2024fairness}
\bibfield{author}{\bibinfo{person}{Yashar Deldjoo}, \bibinfo{person}{Dietmar Jannach}, \bibinfo{person}{Alejandro Bellogin}, \bibinfo{person}{Alessandro Difonzo}, {and} \bibinfo{person}{Dario Zanzonelli}.} \bibinfo{year}{2024}\natexlab{}.
\newblock \showarticletitle{Fairness in Recommender Systems: Research Landscape and Future Directions}.
\newblock \bibinfo{journal}{\emph{User Modeling and User-Adapted Interaction}} \bibinfo{volume}{34}, \bibinfo{number}{1} (\bibinfo{year}{2024}), \bibinfo{pages}{59--108}.
\newblock


\bibitem[Ge et~al\mbox{.}(2022)]%
        {ge2022survey}
\bibfield{author}{\bibinfo{person}{Yingqiang Ge}, \bibinfo{person}{Shuchang Liu}, \bibinfo{person}{Zuohui Fu}, \bibinfo{person}{Juntao Tan}, \bibinfo{person}{Zelong Li}, \bibinfo{person}{Shuyuan Xu}, \bibinfo{person}{Yunqi Li}, \bibinfo{person}{Yikun Xian}, {and} \bibinfo{person}{Yongfeng Zhang}.} \bibinfo{year}{2022}\natexlab{}.
\newblock \showarticletitle{A Survey on Trustworthy Recommender Systems}.
\newblock \bibinfo{journal}{\emph{ACM Transactions on Recommender Systems}} (\bibinfo{year}{2022}).
\newblock


\bibitem[Kazienko and Cambria(2024)]%
        {kazienko2024toward}
\bibfield{author}{\bibinfo{person}{Przemys{\l}aw Kazienko} {and} \bibinfo{person}{Erik Cambria}.} \bibinfo{year}{2024}\natexlab{}.
\newblock \showarticletitle{Toward Responsible Recommender Systems}.
\newblock \bibinfo{journal}{\emph{IEEE Intelligent Systems}} \bibinfo{volume}{39}, \bibinfo{number}{3} (\bibinfo{year}{2024}), \bibinfo{pages}{5--12}.
\newblock


\bibitem[Konstan and Terveen(2021)]%
        {konstan2021human}
\bibfield{author}{\bibinfo{person}{Joseph Konstan} {and} \bibinfo{person}{Loren Terveen}.} \bibinfo{year}{2021}\natexlab{}.
\newblock \showarticletitle{Human-centered Recommender Systems: Origins, Advances, Challenges, and Opportunities}.
\newblock \bibinfo{journal}{\emph{AI Magazine}} \bibinfo{volume}{42}, \bibinfo{number}{3} (\bibinfo{year}{2021}), \bibinfo{pages}{31--42}.
\newblock


\bibitem[Liu et~al\mbox{.}(2024)]%
        {liu2024privacy}
\bibfield{author}{\bibinfo{person}{Yuwen Liu}, \bibinfo{person}{Xiaokang Zhou}, \bibinfo{person}{Huaizhen Kou}, \bibinfo{person}{Yawu Zhao}, \bibinfo{person}{Xiaolong Xu}, \bibinfo{person}{Xuyun Zhang}, {and} \bibinfo{person}{Lianyong Qi}.} \bibinfo{year}{2024}\natexlab{}.
\newblock \showarticletitle{Privacy-preserving Point-of-interest Recommendation Based on Simplified Graph Convolutional Network for Geological Traveling}.
\newblock \bibinfo{journal}{\emph{ACM Transactions on Intelligent Systems and Technology}} \bibinfo{volume}{15}, \bibinfo{number}{4} (\bibinfo{year}{2024}), \bibinfo{pages}{1--17}.
\newblock


\bibitem[Lowdermilk(2013)]%
        {lowdermilk2013user}
\bibfield{author}{\bibinfo{person}{Travis Lowdermilk}.} \bibinfo{year}{2013}\natexlab{}.
\newblock \bibinfo{booktitle}{\emph{User-centered Design: a Developer's Guide to Building User-friendly Applications}}.
\newblock \bibinfo{publisher}{" O'Reilly Media, Inc."}.
\newblock


\bibitem[Shneiderman(2020)]%
        {shneiderman2020human}
\bibfield{author}{\bibinfo{person}{Ben Shneiderman}.} \bibinfo{year}{2020}\natexlab{}.
\newblock \showarticletitle{Human-centered Artificial Intelligence: Reliable, Safe \& Trustworthy}.
\newblock \bibinfo{journal}{\emph{International Journal of Human--Computer Interaction}} \bibinfo{volume}{36}, \bibinfo{number}{6} (\bibinfo{year}{2020}), \bibinfo{pages}{495--504}.
\newblock


\bibitem[Silva et~al\mbox{.}(2024)]%
        {silva2024leveraging}
\bibfield{author}{\bibinfo{person}{{\'I}tallo Silva}, \bibinfo{person}{Leandro Marinho}, \bibinfo{person}{Alan Said}, {and} \bibinfo{person}{Martijn~C Willemsen}.} \bibinfo{year}{2024}\natexlab{}.
\newblock \showarticletitle{Leveraging ChatGPT for Automated Human-centered Explanations in Recommender Systems}. In \bibinfo{booktitle}{\emph{Proceedings of the 29th International Conference on Intelligent User Interfaces}}. \bibinfo{pages}{597--608}.
\newblock


\bibitem[Wang et~al\mbox{.}(2024)]%
        {wang2024trustworthy}
\bibfield{author}{\bibinfo{person}{Shoujin Wang}, \bibinfo{person}{Xiuzhen Zhang}, \bibinfo{person}{Yan Wang}, {and} \bibinfo{person}{Francesco Ricci}.} \bibinfo{year}{2024}\natexlab{}.
\newblock \showarticletitle{Trustworthy Recommender Systems}.
\newblock \bibinfo{journal}{\emph{ACM Transactions on Intelligent Systems and Technology}} \bibinfo{volume}{15}, \bibinfo{number}{4} (\bibinfo{year}{2024}), \bibinfo{pages}{1--20}.
\newblock


\bibitem[Wu et~al\mbox{.}(2024)]%
        {wu2024accelerating}
\bibfield{author}{\bibinfo{person}{Yunfan Wu}, \bibinfo{person}{Qi Cao}, \bibinfo{person}{Shuchang Tao}, \bibinfo{person}{Kaike Zhang}, \bibinfo{person}{Fei Sun}, {and} \bibinfo{person}{Huawei Shen}.} \bibinfo{year}{2024}\natexlab{}.
\newblock \showarticletitle{Accelerating the Surrogate Retraining for Poisoning Attacks against Recommender Systems}. In \bibinfo{booktitle}{\emph{Proceedings of the 18th ACM Conference on Recommender Systems}}. \bibinfo{pages}{701--711}.
\newblock


\bibitem[Yoo et~al\mbox{.}(2024)]%
        {yoo2024ensuring}
\bibfield{author}{\bibinfo{person}{Hyunsik Yoo}, \bibinfo{person}{Zhichen Zeng}, \bibinfo{person}{Jian Kang}, \bibinfo{person}{Ruizhong Qiu}, \bibinfo{person}{David Zhou}, \bibinfo{person}{Zhining Liu}, \bibinfo{person}{Fei Wang}, \bibinfo{person}{Charlie Xu}, \bibinfo{person}{Eunice Chan}, {and} \bibinfo{person}{Hanghang Tong}.} \bibinfo{year}{2024}\natexlab{}.
\newblock \showarticletitle{Ensuring User-side Fairness in Dynamic Recommender Systems}. In \bibinfo{booktitle}{\emph{Proceedings of the ACM on Web Conference 2024}}. \bibinfo{pages}{3667--3678}.
\newblock


\bibitem[Zhang et~al\mbox{.}(2023)]%
        {zhang2023robust}
\bibfield{author}{\bibinfo{person}{Kaike Zhang}, \bibinfo{person}{Qi Cao}, \bibinfo{person}{Fei Sun}, \bibinfo{person}{Yunfan Wu}, \bibinfo{person}{Shuchang Tao}, \bibinfo{person}{Huawei Shen}, {and} \bibinfo{person}{Xueqi Cheng}.} \bibinfo{year}{2023}\natexlab{}.
\newblock \showarticletitle{Robust Recommender System: a Survey and Future Directions}.
\newblock \bibinfo{journal}{\emph{arXiv preprint arXiv:2309.02057}} (\bibinfo{year}{2023}).
\newblock


\bibitem[Zhang et~al\mbox{.}(2024a)]%
        {zhang2024improving}
\bibfield{author}{\bibinfo{person}{Kaike Zhang}, \bibinfo{person}{Qi Cao}, \bibinfo{person}{Yunfan Wu}, \bibinfo{person}{Fei Sun}, \bibinfo{person}{Huawei Shen}, {and} \bibinfo{person}{Xueqi Cheng}.} \bibinfo{year}{2024}\natexlab{a}.
\newblock \showarticletitle{Improving the Shortest Plank: Vulnerability-Aware Adversarial Training for Robust Recommender System}. In \bibinfo{booktitle}{\emph{Proceedings of the 18th ACM Conference on Recommender Systems}}. \bibinfo{pages}{680--689}.
\newblock


\bibitem[Zhang et~al\mbox{.}(2024b)]%
        {zhang2024lorec}
\bibfield{author}{\bibinfo{person}{Kaike Zhang}, \bibinfo{person}{Qi Cao}, \bibinfo{person}{Yunfan Wu}, \bibinfo{person}{Fei Sun}, \bibinfo{person}{Huawei Shen}, {and} \bibinfo{person}{Xueqi Cheng}.} \bibinfo{year}{2024}\natexlab{b}.
\newblock \showarticletitle{LoRec: Combating Poisons with Large Language Model for Robust Sequential Recommendation}. In \bibinfo{booktitle}{\emph{Proceedings of the 47th International ACM SIGIR Conference on Research and Development in Information Retrieval}}. \bibinfo{pages}{1733--1742}.
\newblock


\bibitem[Zhang et~al\mbox{.}(2024c)]%
        {zhang2024understanding}
\bibfield{author}{\bibinfo{person}{Kaike Zhang}, \bibinfo{person}{Qi Cao}, \bibinfo{person}{Yunfan Wu}, \bibinfo{person}{Fei Sun}, \bibinfo{person}{Huawei Shen}, {and} \bibinfo{person}{Xueqi Cheng}.} \bibinfo{year}{2024}\natexlab{c}.
\newblock \showarticletitle{Understanding and Improving Adversarial Collaborative Filtering for Robust Recommendation}.
\newblock \bibinfo{journal}{\emph{arXiv preprint arXiv:2410.22844}} (\bibinfo{year}{2024}).
\newblock


\bibitem[Zhou et~al\mbox{.}(2024)]%
        {zhou2024bee}
\bibfield{author}{\bibinfo{person}{Xiaofei Zhou}, \bibinfo{person}{Yushan Zhou}, \bibinfo{person}{Yunfan Gong}, \bibinfo{person}{Zhenyao Cai}, \bibinfo{person}{Annie Qiu}, \bibinfo{person}{Qinqin Xiao}, \bibinfo{person}{Alissa~N Antle}, {and} \bibinfo{person}{Zhen Bai}.} \bibinfo{year}{2024}\natexlab{}.
\newblock \showarticletitle{" Bee and I need diversity!" Break Filter Bubbles in Recommendation Systems through Embodied AI Learning}. In \bibinfo{booktitle}{\emph{Proceedings of the 23rd Annual ACM Interaction Design and Children Conference}}. \bibinfo{pages}{44--61}.
\newblock


\end{thebibliography}
